\documentclass{sf2a-conf2014}
\usepackage{graphicx}
\usepackage{hyperref}
\usepackage[]{natbib}  
\usepackage{epstopdf}

\def\BibTeX{{\rm B\kern-.05em{\sc i\kern-.025em b}\kern-.08em
    T\kern-.1667em\lower.7ex\hbox{E}\kern-.125emX}}
\bibpunct{(}{)}{;}{a}{}{,}  

\begin{document}

\TitreGlobal{SF2A 2014}


\title{The influence of the magnetic topology on the wind braking of sun-like stars.}

\runningtitle{}

\author{V. R\'eville}\address{Laboratoire AIM, DSM/IRFU/SAp, CEA Saclay, 91191 Gif-sur-Yvette Cedex}

\author{A. S. Brun$^1$}
\author{S. P. Matt}\address{Department of Physics and Astronomy, University of Exeter, Stocker Road, Exeter EX4 4SB, UK}

\author{A. Strugarek$^1,$}\address{D\'epartement de physique, Universit\'e de Montr\'eal, C.P. 6128 Succ. Centre-Ville, Montr\'eal, QC H3C-3J7, Canada}

\author{R. Pinto$^1,$}\address{LESIA, Observatoire de Paris-Meudon, 5 place Jules Janssen, 92195 MEUDON Cedex}




\setcounter{page}{237}


\maketitle


\begin{abstract}
Stellar winds are thought to be the main process responsible for the spin down of main-sequence stars. The extraction of angular momentum by a magnetized wind has been studied for decades, leading to several formulations for the resulting torque. However, previous studies generally consider simple dipole or split monopole stellar magnetic topologies. Here we consider in addition to a dipolar stellar magnetic field, both quadrupolar and octupolar configurations, while also varying the rotation rate and the magnetic field strength. 60 simulations made with a 2.5D, cylindrical and axisymmetric set-up and computed with the PLUTO code were used to find torque formulations for each topology. We further succeed to give a unique law that fits the data for every topology by formulating the torque in terms of the amount of open magnetic flux in the wind. We also show that our formulation can be applied to even more realistic magnetic topologies, with examples of the Sun in its minimum and maximum phase as observed at the Wilcox Solar Observatory, and of a young K-star (TYC-0486-4943-1) whose topology has been obtained by Zeeman-Doppler Imaging (ZDI).
\end{abstract}



\section{Introduction}
To explain the observed pressure ratio of order $10^{14}$ between the base of the solar corona and the interstellar medium, \citet{Parker1958} introduced the idea of a non-hydrostatic expanding solar atmosphere, the solar wind. This accelerated outflow has been observed since with typical speed between 400 km/s and 1000 km/s around the Earth. This process is thought to occur in all cool stars with an upper convective layer, \textit{i.e.} from stellar type M to F, or $0.5 M_{\odot}$ to $1.4 M_{\odot}$. \citet{Schatzman1962}, \citet{Parker1963}, \citet{WeberDavis1967} and \citet{Mestel1968} then introduced the effect of both magnetic field and rotation thus creating the magnetic rotator theory which is now the standard MHD theory for stellar winds and the main process responsible for the braking of solar-like stars observed all along the main sequence \citep{IrwinBouvier2009}. It combines the driving of the wind due to the pressure gradient and the magneto-centrifugal effect. \citet{WeberDavis1967} used a simple one dimensional model (at the equator) to quantify the angular momentum carried by the plasma and demonstrated that the torque $\tau_w$ can be expressed:

\begin{equation}
\tau_w = \dot{M}_w \Omega_* R_A^2,
\label{torque}
\end{equation}

where $\dot{M}_w$ is the integrated mass loss rate,  $\Omega_*$ the rotation rate of the star and $R_A$ the Alfv\'en radius, \textit{i.e.} the radius at which the velocity field reach the Alfv\'en speed $v_A=B_p/\sqrt{4 \pi \rho}$ ($B_p$ is the poloidal component of the magnetic field in this model).

In order to find a formulation for a realistic multi-dimensional outflow, we define an average value for the Alfv\'en radius (which is the cylindrical radius, the distance from the rotation axis) such that equation (\ref{torque}) is always true: 

\begin{equation}
\langle R_A \rangle  = \sqrt{\frac{\tau_w}{\dot{M}_w \Omega_*}}
\label{AlfRad}
\end{equation}

\section{Numerical setup}

\begin{figure}[h!]
\center
\begin{tabular}{ccc}
\includegraphics[scale=0.6]{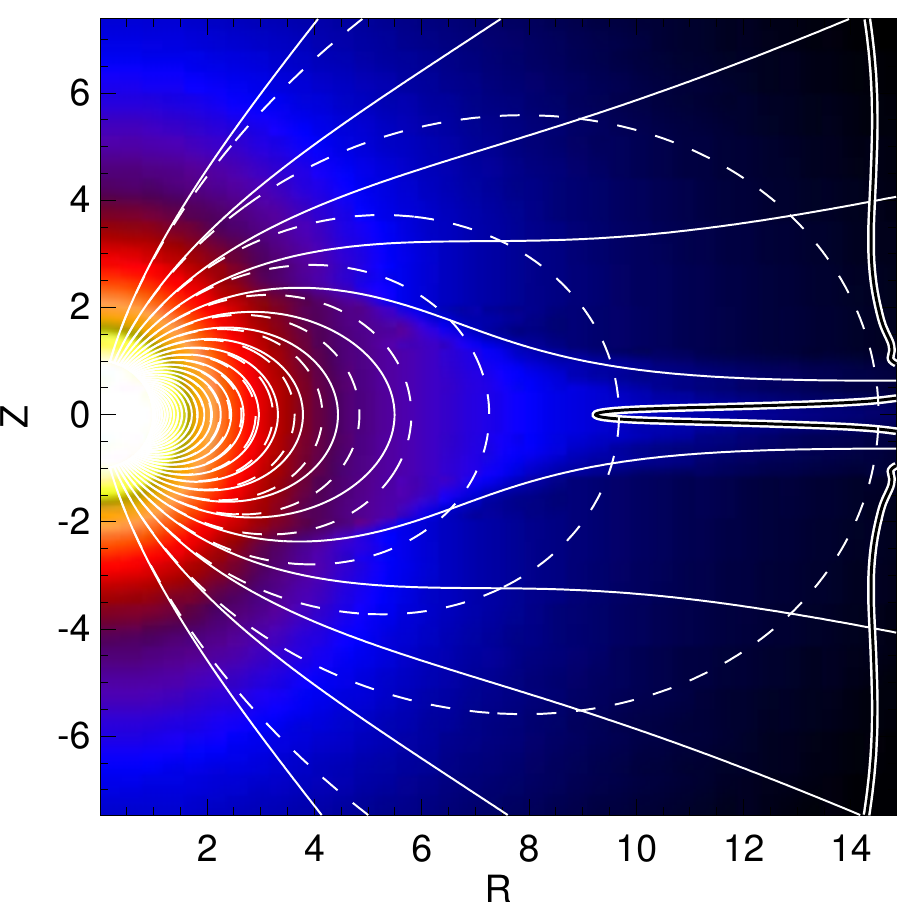} & \includegraphics[scale=0.6]{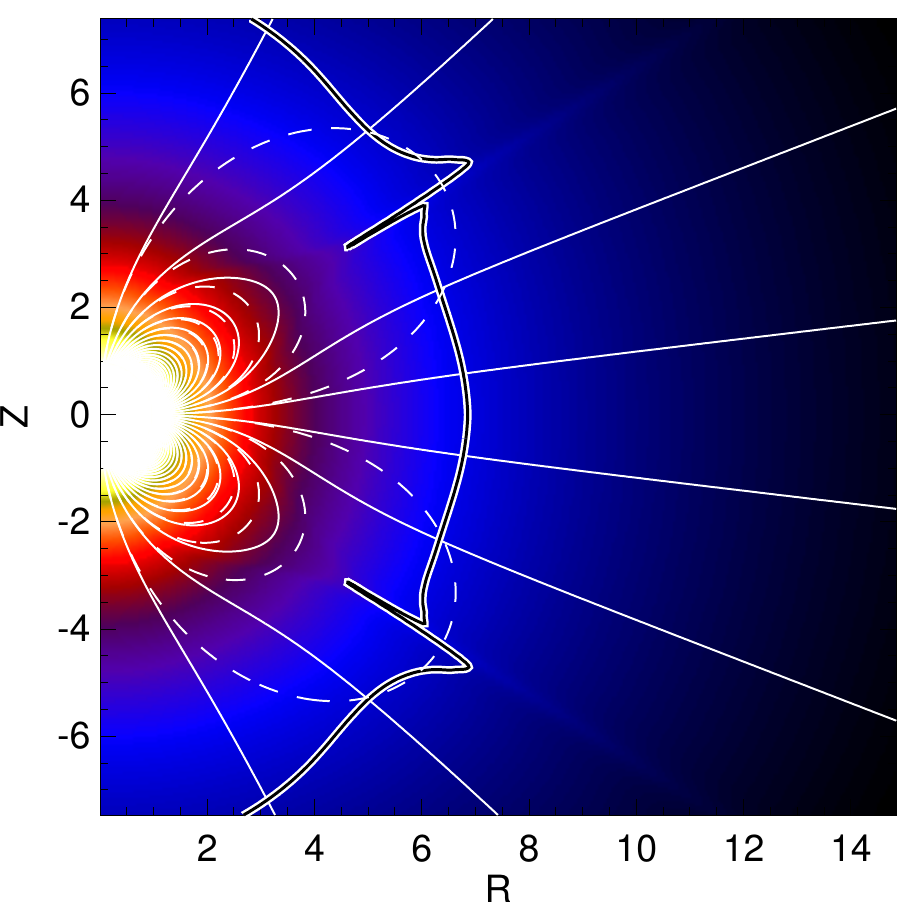} & \includegraphics[scale=0.6]{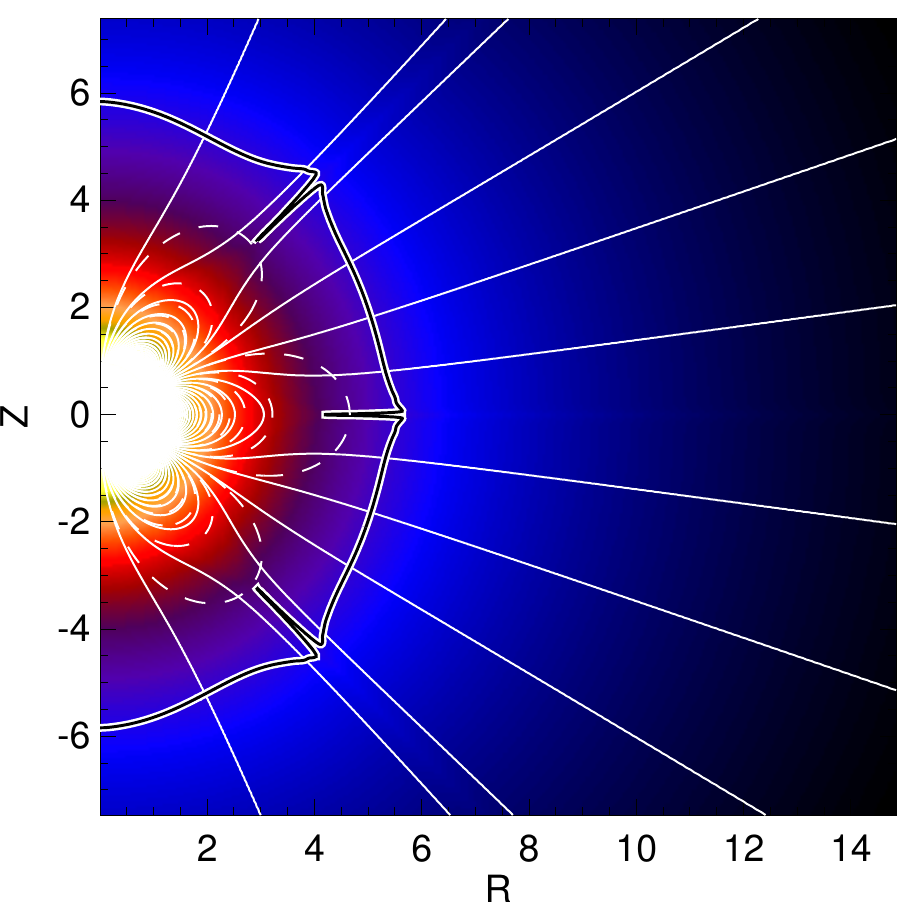}\\
\end{tabular}
\includegraphics[scale=0.25]{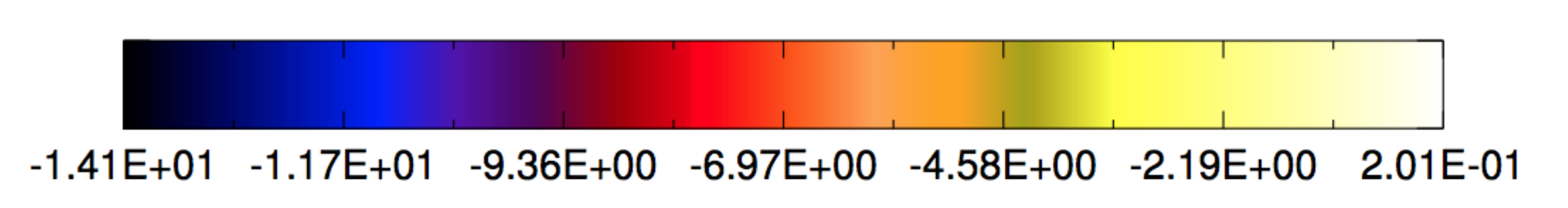}
\caption{Magnetic field lines of initial topologies (dashed lines) and final state (continuous lines) for a typical case (case 2 of \citet{Reville2014}) and the three topologies. The color background is the logarithm of density.}
\label{InitTopo}
\end{figure}

In our study we use the compressible MHD code PLUTO, to perform 2.5D simulations, similarly to \citet{Matt2012} and \citet{MP2008}. We initialize the star thanks to a one dimensional Parker solution for the outflow and let the code evolve the equations of ideal MHD. The magnetic field is initially a pure dipole, quadrupole, octupole or the sum of the three components (see section \ref{Realistic}). The outflow then non-linearly interacts with the magnetic field, opens part of the field lines and reach a steady-state. In Figure \ref{InitTopo} we show the three different magnetic topologies in the initial and final states. For a precise description of the boundary conditions, see \citet{Strugarek2014IAU} and \citet{Reville2014}. We used 20 cases of \citet{Matt2012}, that we ran with the three topologies. 

\section{Braking Laws}

We fit our set of simulations with the formulation of \citet{Matt2012}:

\begin{equation}
\frac{\langle R_A \rangle}{R_*}=K_1 [\frac{\Upsilon}{(1+f^2/K_2^2)^{1/2}}]^m \; \; \; {\rm where} \; \; \; \Upsilon \equiv \frac{B_*^2 R_*^2}{\dot{M}_w v_{esc}}
\label{form1}
\end{equation}

is the magnetization parameter introduced in \citet{MP2008}. A similar parameter has been introduced before in \citet{UdDoula2002}), where the terminal velocity $v_{\infty}$ was used instead of the escape velocity $v_{esc} \equiv \sqrt{(2 G M_*)/R_*}$. Both characterize the magnetization of the wind, which is the ratio of the magnetic field energy and the kinetic energy of the wind. $B_*$ is the magnetic field strength taken at the surface and at the equator of the star, $\dot{M}_w$ is the integrated mass-loss rate and $f$ is the fraction of break-up rate, \textit{i.e.} the ratio between the rotation rate at the equator of the star (in our simulations the star has a solid body rotation) and the keplerian speed that is defined by:

\begin{equation}
f \equiv \Omega_* R_*^{3/2}(GM_*)^{-1/2}.
\end{equation}


\begin{table}
\center
\begin{tabular}{lccc}
\hline
\hline
Topology & $K_1$ & $K_2$ & $m$ \\
 \hline
Dipole  & $2.0 \pm 0.1$ & $0.2 \pm 0.1$ & $0.235 \pm 0.007$ \\
Quadrupole & $1.7 \pm 0.3$ & $0.2 \pm 0.1$ & $0.15 \pm 0.02$ \\
Octupole  & $1.7 \pm 0.3$ & $0.2 \pm 0.1$ & $0.11\pm 0.02$ \\
 \hline
 \vspace{-0.1cm} & & \\ 
 & $K_3$ &$K_4$& $m$ \\
\hline 
\vspace{-0.1cm} & &\\ 
Topology independent & $1.4 \pm 0.1$ & $0.06 \pm 0.01$ & $0.31 \pm 0.02$ \\
\hline
\end{tabular}
\caption{Parameters of the fit to equation \ref{form1} made independently for each topology. The values $K_1$,$K_2$ and $m$ corresponds to formulation \ref{form1} for the dipolar, quadrupolar and octupolar configurations. The parameter $K_3$ and $K_4$ for the topology independent formulation \ref{form2} are given as well.}
\label{fitpar}
\end{table}

Table \ref{fitpar} shows the different parameters of the fit depending on the topology. The influence of topology can mainly be seen through the change in the power-law exponent $m$. It decreases with higher order multipole. This is due to the radial dependency of the magnetic field (see \citealt{Kawaler1988,Reville2014}). Hence the braking is more efficient for lower order multipole for a given magnetic field strength. As shown in \citet{MP2008} and \citet{Matt2012}, increasing the magnetic field strength increases the average Alfv\'en radius (and thus the torque) while increasing the rotation rate makes the average Alfv\'en radius smaller. Rotation adds acceleration through magneto-centrifugal forces \citep{Ustyugova1999,Reville2014}.

In order to obtain a topology independent formulation for the magnetic torque we define a new magnetization parameter using the open flux:

\begin{equation}
\Upsilon_{open} \equiv \frac{\Phi_{open}^2}{R_*^2 \dot{M}_w v_{esc}}, 
\end{equation}

and consider a new formulation of $\langle R_A \rangle$ as a function of $\Upsilon_{open}$:

\begin{equation}
\frac{\langle R_A \rangle}{R_*}=K_3 [\frac{\Upsilon_{open}}{(1+f^2/K_4^2)^{1/2}}]^m 
\label{form2}
\end{equation}

\begin{figure}[h!]
\center
\begin{tabular}{cc}
\includegraphics[scale=0.5]{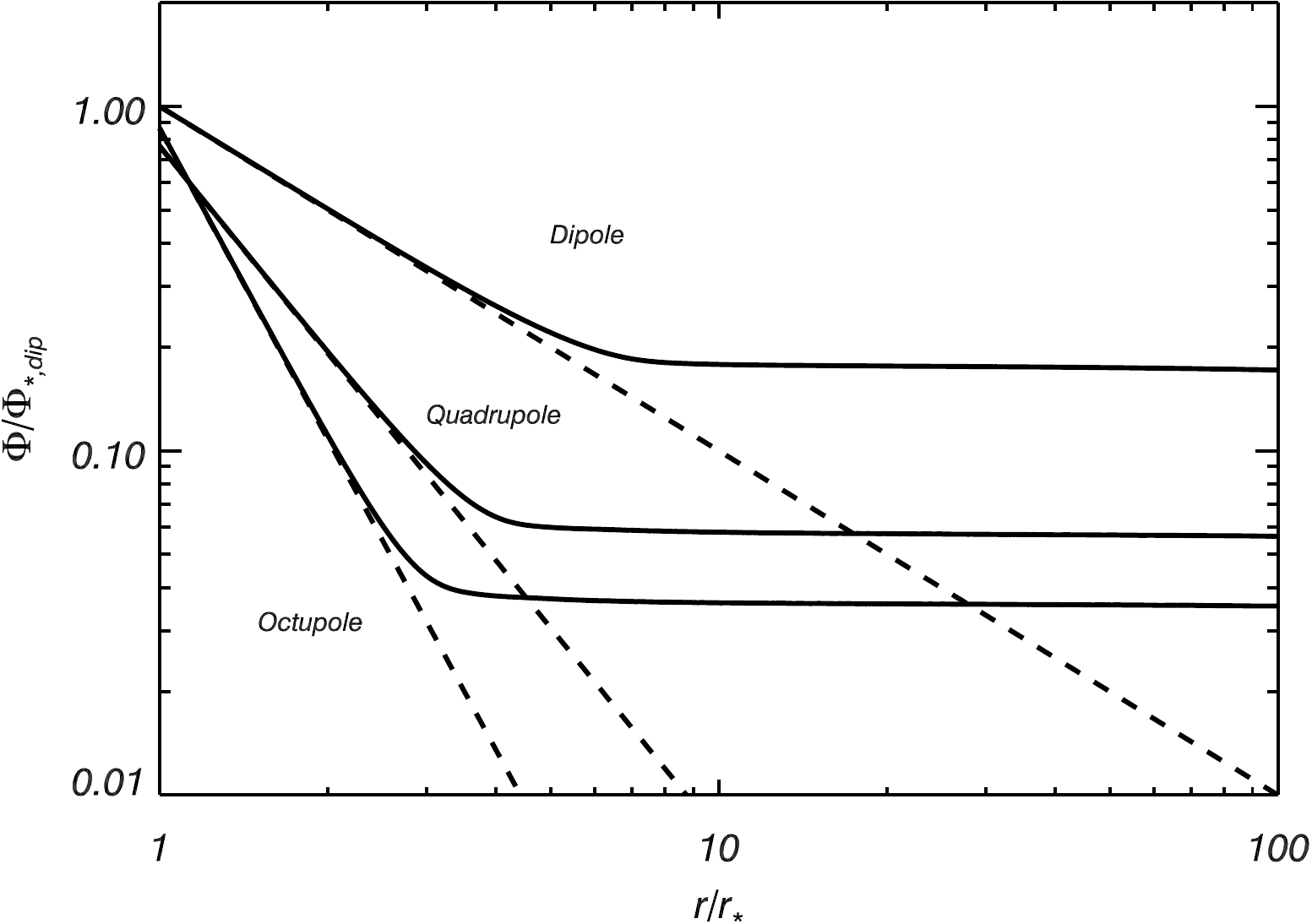} &\includegraphics[scale=0.5]{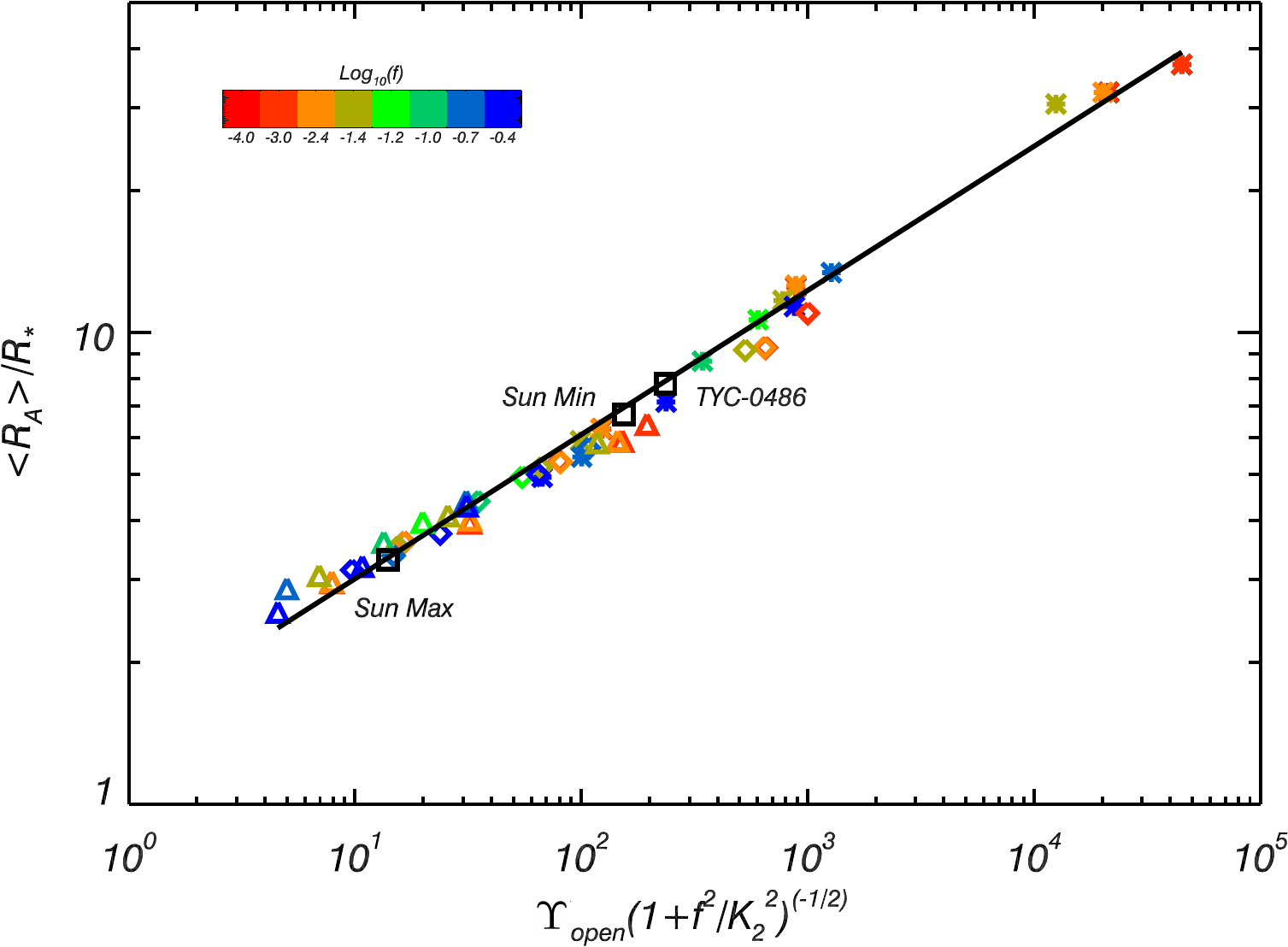}\\
\end{tabular}
\caption{The integrated unsigned magnetic fluxes for the three topologies are shown on the left panel. The one law fit for the three topologies is shown on the right panel. Colors represent the rotation rate, from red to blue as rotation increases. Symbols stand for the topology, stars for dipole, diamonds for quadrupole, triangles for octupole. The influence of the rotation is captured in x-axis variable. Black squares are complex topologies (Sun Min, Sun Max and TYC-0486-4943-1 introduced in section \ref{Realistic}), which demonstrates that winds with combination of magnetic multipoles follow the same braking law as those with single-mode topologies.}
\label{IndForm}
\end{figure}

Figure \ref{IndForm} shows on the left panel the unsigned magnetic flux integrated over concentric spheres for the three different topologies. Beyond a few stellar radii, all the field lines are open and the magnetic flux becomes constant. This constant value is the open flux $\Phi_{open}$ used in our new formulation. On the right panel of Figure \ref{IndForm}, we show that a single braking law can be used to fit our 60 simulations. The parameter of the fit are given in Table \ref{fitpar}.

\section{Realistic Topologies}
\label{Realistic}

\begin{figure}
\center
\begin{tabular}{ccc}
\includegraphics[scale=0.55]{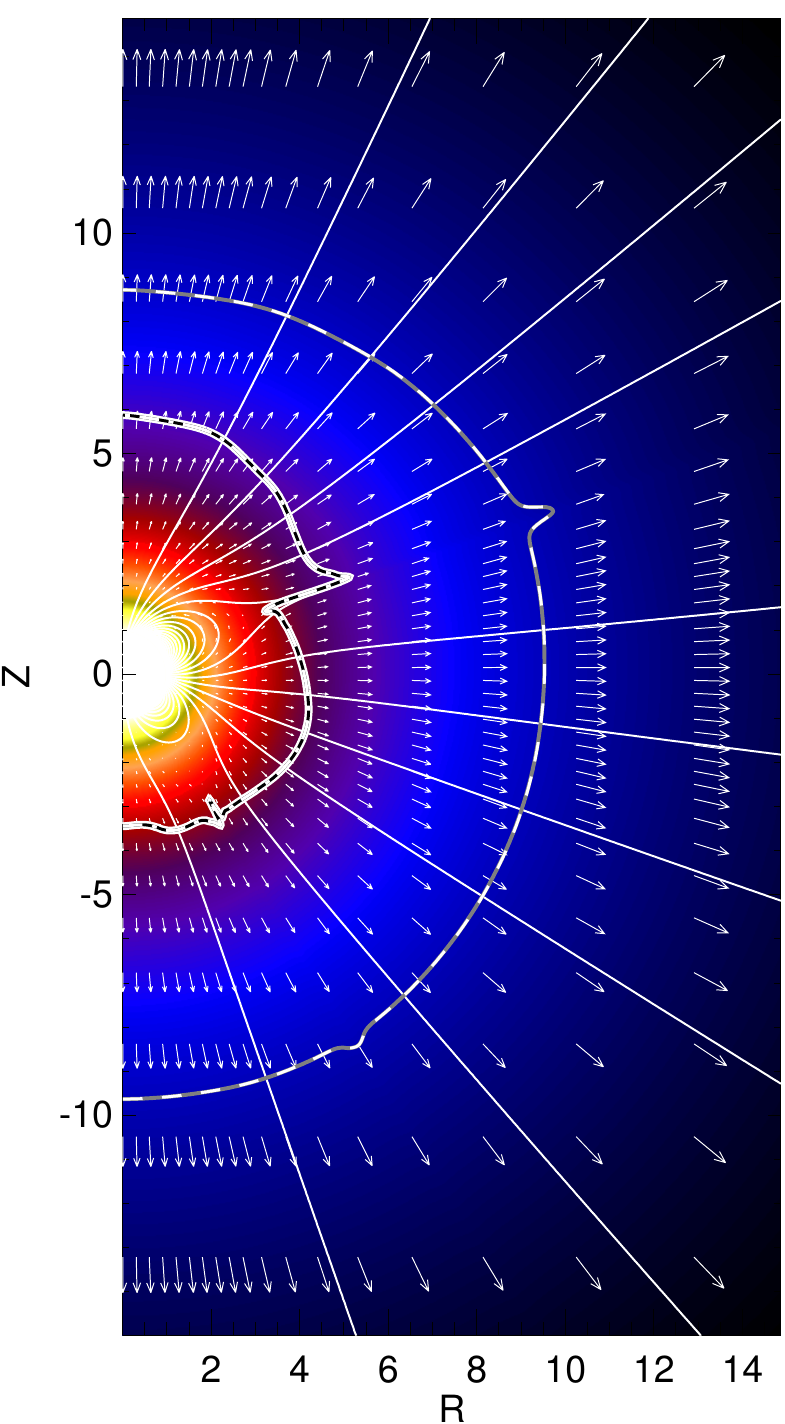} & \includegraphics[scale=0.55]{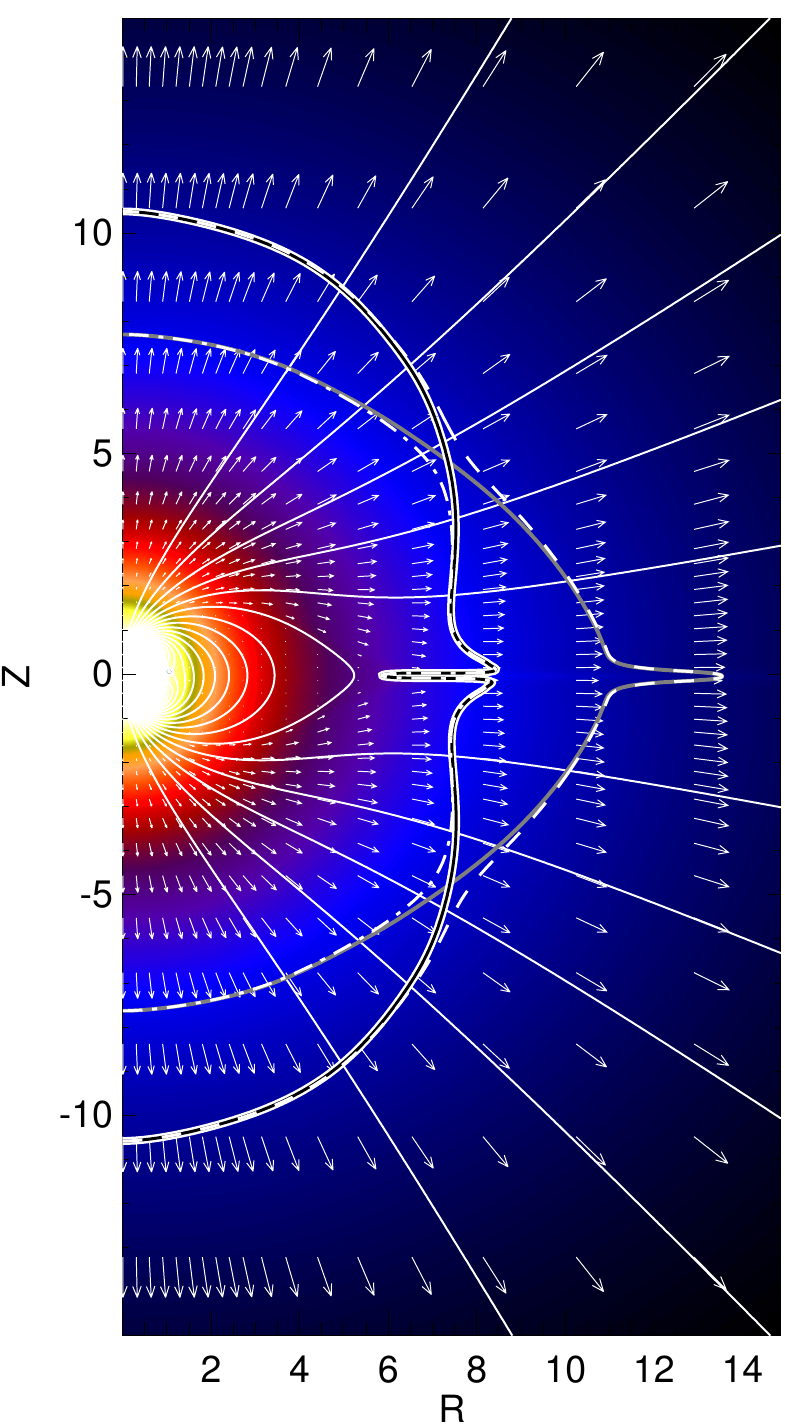} &\includegraphics[scale=0.59]{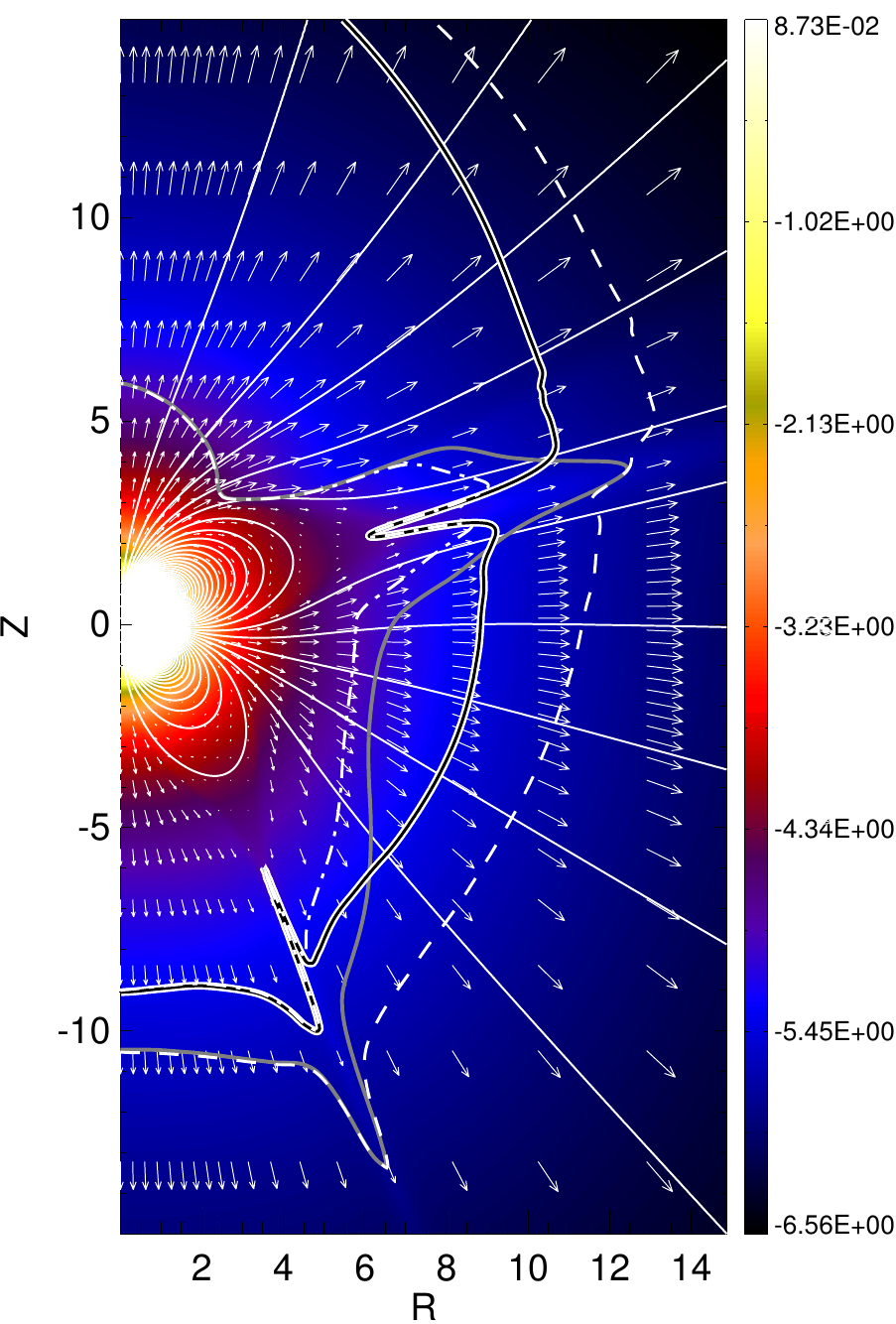} \\
\textbf{Solar Maximum} & \textbf{Solar Minimum}& \textbf{TYC-0486-4943-1}\\
\end{tabular}
\caption{Steady state solutions of three winds with a realistic magnetic topology extracted from Wilcox Solar Observatory data \citep{DeRosa2012} for the solar cases and from a ZDI Map of TYC-0486-4943-1. Only the axisymmetric component till $l=3$ are taken into account. The wind Alfv\'en surfaces are shown in the same format as Figure \ref{InitTopo}. The background is the logarithm of the density and we added velocity arrows in white.}
\label{RealSol}
\end{figure}

Our topology independent formulation is able to fit three different multipoles. However magnetic field topologies of stars are a combination of those three modes and higher order multipoles. We then tried to compare this formulation with the resulting torque of actual stars whose magnetic field is approximated as the sum of a dipolar, quadrupolar and octupolar components. We took the axisymmetric components of these three modes of the Sun during the minimum and the maximum of activity of the cycle 22. The spherical harmonics decomposition have been obtained from magnetograms of the Wilcox Observatory \citep[see][]{DeRosa2012}. The simulations gave us a torque that is predicted by our formulation as shown in Figure \ref{IndForm}. We also used the spherical harmonics decomposition of the surface magnetic field of TYC-0486 obtained through Zeeman-Doppler-Imaging, and performed a simulation of the wind of this young K-Star (see Figure \ref{RealSol}). The torque is again well predicted by our formulation even though we assume that the coronal temperature of this star is equal to the Sun's. Changing the temperature might change the braking law's coefficients $K_3$ and $K_4$ slightly but we expect $m$ to be robust. A more systematic study is needed.

\section{Open Flux Calculations and Perspectives}

With our topology independent formulation, the stellar wind magnetic torque can  be written:

\begin{equation}
\tau_w= \dot{M}_w^{1-2m} \Omega_* R_*^{2-4m} K_3^2 (\frac{\Phi_{open}^2}{v_{esc}(1+f^2/K_4^2)^{1/2}})^{2m}.
\end{equation}

However the prediction of the open magnetic flux from the surface magnetic field is not a trivial task without running a MHD simulation. Some correlation with the surface flux have been proposed \citep{Vidotto2014}, but they rely on heavy 3D simulations that we eventually want to avoid to provide observers with efficient tools to compute the magnetic torque.

\begin{figure}
\center
\begin{tabular}{cc}
\includegraphics[scale=0.7]{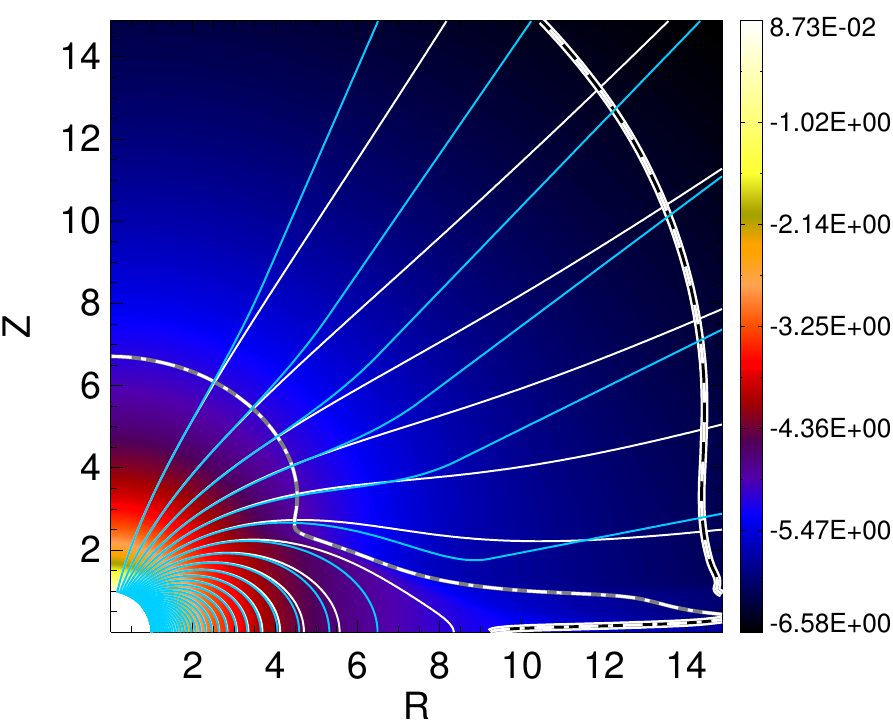} & \includegraphics[scale=0.7]{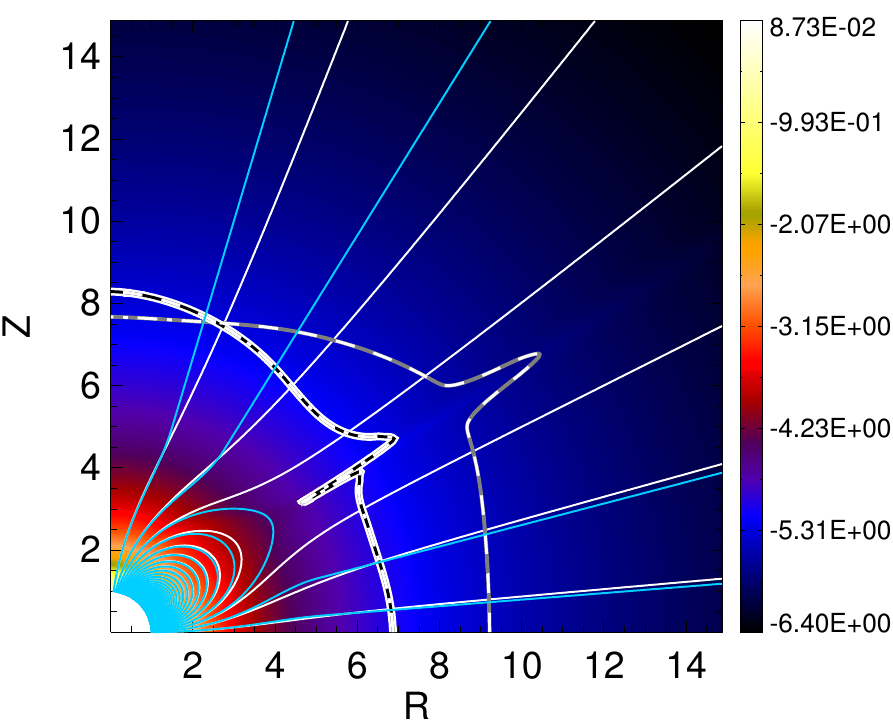}\\
\end{tabular}
\caption{Comparison of the field lines obtained with the wind simulation (in white) and a potential extrapolation at the optimum source surface (in cyan). In both cases (dipolar and quadrupolar), the magnetic loops are well reproduced.}
\label{Dip2RSS}
\end{figure}

Considering a potential extrapolation of the surface magnetic field \citep{SchrijverDeRosa2003}, the location of the source surface is the only thing needed to compute the open flux. And such a reconstruction is by far easier to compute than a whole wind simulation. From our set of simulations we developed a method that minimizes the difference (through a gradient descent) between the potential extrapolation and the steady-state wind solution, and find an optimum value for the source surface. The comparison is shown in Figure \ref{Dip2RSS}, the white field lines are from our simulation while the cyan field lines are drawn from a potential that has been reconstructed using the optimum source surface we computed. Closed magnetic loops are well reproduced by this model. However the potential field extrapolation provides a magnetic field that is completely radial by definition beyond the source surface, whereas it is not exactly the case in our simulation, especially for highly rotating cases. But this has no influence on the value of the open flux, since the only thing that matters is how much magnetic flux is confined in the magnetic loops. We found that a good proxy for the location of the source surface radius is the average radius where the ram pressure plus the thermal pressure become higher than the magnetic pressure. The agreement for this average radius with the optimum source surface is within 10\% for our slowly rotating dipolar cases. Taking into account rotation and topology might introduce some corrections, and more investigations are needed. They will be reported in a subsequent paper where we will provide tools to determine the source surface as a function of the stellar parameters.

Consequently, with our topology independent formulation and the open flux computed from ZDI Maps, we are able to give precise estimates of the torque exerted on the star, if the mass loss rate have been prescribed through analytical models \citep{CranmerSaar2011} or observations \citep{WoodRev2004}. Interestingly, this formulation could also provide constraints on the mass loss rate of clusters whose braking time scales are known.

\begin{acknowledgements}
We would like to thank Colin Folsom, Pascal Petit for the magnetic field decomposition coefficients of TYC- 0486-4943-1, J\'erome Bouvier and the ANR TOUPIES project which aim to understand the evolution of star?s spin rates, the ERC STARS2 (www.stars2.eu) and CNES support via our Solar Orbiter funding.
\end{acknowledgements}


\bibliographystyle{aa}  
\bibliography{biblio}

\end{document}